\documentclass[letterpaper, 10 pt, conference]{ieeeconf}
\IEEEoverridecommandlockouts    
\usepackage{times}

\usepackage{multicol}
\usepackage[bookmarks=true]{hyperref}

\usepackage{amsfonts}       
\usepackage{amsmath}
\usepackage{amssymb}
\usepackage{graphicx}
\usepackage{epstopdf}
\usepackage{mathtools}
\usepackage{dsfont}
\usepackage{tikz}
\usepackage{siunitx}
\usepackage{hyperref}
\usepackage[font=footnotesize]{caption}
\usepackage[noabbrev,capitalize]{cleveref}
\usepackage{balance}    
\usepackage{xcolor}
\usepackage{tcolorbox}
\usepackage{algorithm}
\usepackage{algpseudocode}
\usepackage{mathtools}
\usepackage{dsfont}
\usepackage{mathrsfs}
\usepackage{stmaryrd}
\usepackage{placeins}
\usepackage{upgreek}
\usepackage{wrapfig}
\usepackage{bm}
\usepackage{cases}
\usepackage{bbm}
\usepackage{noel}

\title{\LARGE \bf Constructive Nonlinear Control of Underactuated Systems \\ via Zero Dynamics Policies}

\begin{document}
\author{William Compton$^*$, Ivan Dario Jimenez Rodriguez$^*$, Noel Csomay-Shanklin$^*$, \\Yisong Yue, Aaron D. Ames%
\thanks{$^*$denotes equal contribution. Authors are with the Department of Computing and Mathematical Sciences, California Institute of Technology,
Pasadena, CA 91125. This research is supported by Technology Innovation Institute (TII), NSF Graduate Research Fellowship No. DGE‐1745301, NSF CPS Grant \#1918655, JPL RTD 1643049.}
}

\maketitle

\vspace{-6mm}
\begin{abstract}%
Stabilizing underactuated systems is an inherently challenging control task due to fundamental limitations on how the control input affects the unactuated dynamics.
Decomposing the system into actuated (output) and unactuated (zero) coordinates provides useful insight as to how input enters the system dynamics.
In this work, we leverage the structure of this decomposition to formalize the idea of \emph{Zero Dynamics Policies (ZDPs)}---a mapping from the unactuated coordinates to desired actuated coordinates.
Specifically, we show that a ZDP exists in a neighborhood of the origin, and prove that combining output stabilization with a ZDP results in stability of the full system state. 
We detail a constructive method of obtaining ZDPs in a neighborhood of the origin, and propose a learning-based approach which leverages optimal control to obtain ZDPs with much larger regions of attraction. We demonstrate that such a paradigm can be used to stabilize the canonical underactuated system of the cartpole, and showcase an improvement over the nominal performance of LQR.
\end{abstract}

\IEEEpeerreviewmaketitle



\section{Introduction}

\textit{Underactuated} systems -- e.g. legged robots, dexterous manipulators, and systems with strict actuator limits -- are inherently challenging to control due to the present passive dynamics.
Specifically, one cannot directly actuate all degrees of freedom, which results in dynamics that cannot be arbitrarily shaped.
In the case when these passive dynamics are stable, constructive feedback controllers can be synthesized, e.g., input-output linearization \cite{sastry_linearization_1999} and control Lyapunov functions \cite{sontag_control-lyapunov_1999}. 
Yet stabilizing underactuated systems without stability assumptions on the passive dynamics remains a challenging problem. 
To address this problem, this paper builds upon the observation that the existence of passive (not directly actuated) dynamics does not necessarily imply a loss of stabilizability---rather achieving stable behaviors requires careful coordination of the actuated degrees of freedom.  

To stabilize underactuated systems, we leverage \textit{Zero dynamics}: a powerful tool for analyzing underactuated systems which considers the residual dynamics when all outputs have been zeroed \cite{isidori_elementary_1995}.
Traditional Zero Dynamics approaches craft output coordinates whose zero dynamics are then algebraically checked for stability.
Underactuation compels this guess-and-check approach due to the challenges associated with finding
feedback linearizable output coordinates which span the complete state space \cite{bedrossian_nonlinear_nodate}.
This paper takes a different approach, wherein we view the outputs as learnable design variables, enabling stable zero dynamics by construction.  
We prove that stabilizing to the zero dynamics surface defined by the learned outputs results in stability of the overall system.

\begin{figure}[t!]
    \centering
    \includegraphics[width=\columnwidth]{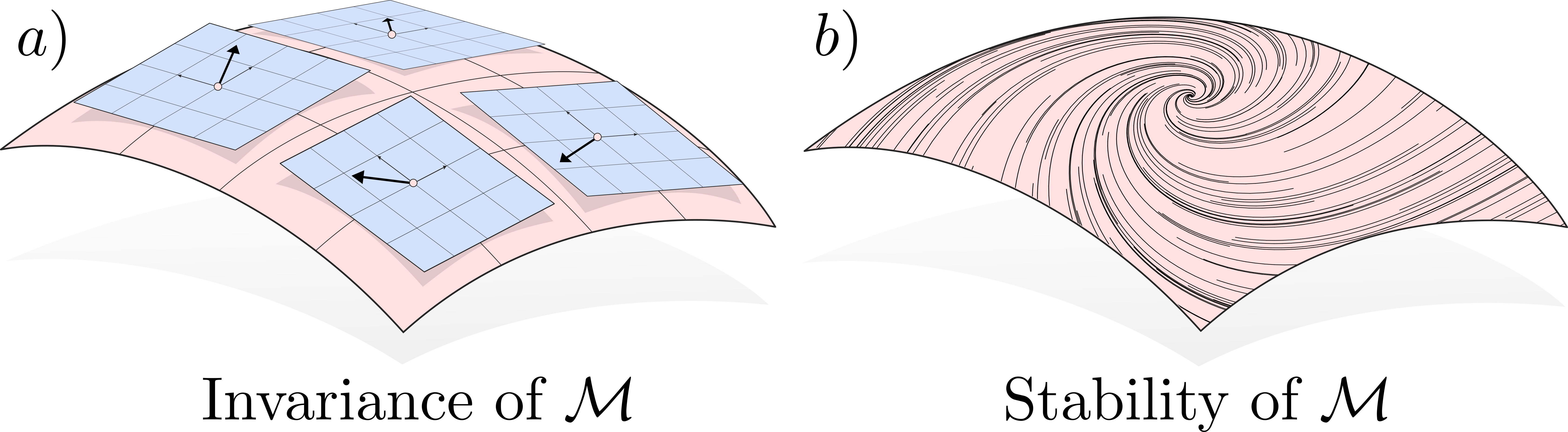}
    \caption{The two conditions required of the zeroing manifold: a) controlled invariance, and b) stable zero dynamics.}
    \label{fig:1}
    \vspace{-6mm}
\end{figure}

The perspective taken in this paper has origins in the stabilization of non-minimum phase systems \cite{devasia_nonlinear_1996, tomlin_output_1995, isidori_output_1990}, and extends work proposing the design of a stabilizing zero dynamics manifold in the context of Hybrid Zero Dynamics (HZD) for bipedal walking \cite{westervelt_hybrid_2003}. 
These methods, however, are often domain-specific and challenging to synthesize, leading many practitioners to turn to optimal control techniques instead.
Both approximate value function feedback \cite{miki2022learning,siekmann2021blind} and receding horizon Model Predictive Control (MPC) are common variations of this approach \cite{csomay2023nonlinear,grandia2023perceptive}. 
Although practically useful, these methods provide limited insight as to why stabilizing underactuated systems is fundamentally difficult.
To address this issue, we take inspiration from legged locomotion \cite{kuindersma2016optimization,full1999templates,raibert1984hopping}, where mappings between the underactuated coordinates to the actuated coordinates are central to controller synthesis.  
The authors exploited this connection, in conjunction with HZD, in \cite{rodriguez_neural_2022} where planar biped walking was generated by enforcing barrier function certificates on the zero dynamics manifold through the use of learned output parameters.   
Our goal in this work is to formalize and unify the above approaches. 

This paper presents a framework for the constructive stabilization of underactuated systems through \emph{Zero Dynamics Policies (ZDPs)}. 
These policies take the form of a mapping from the underactuated states to the actuated states, which defines a controlled invariant and stable manifold, represented in \cref{fig:1}. Stabilizing this manifold via output tracking results in guaranteed stability of the overall system. 
Furthermore, we prove that for (locally) controllable nonlinear systems such a mapping exists and can be constructed analytically in a neighborhood of the origin.
Finally, we leverage optimal control and machine learning as a constructive method for extending the region of validity of these ZDPs. 
We demonstrate that the use of ZDPs leads to a larger region of attraction than traditional control methods on the classic cartpole system. 

\section{Preliminaries}

Consider a control affine nonlinear system
\begin{align}\label{eqn:x_affine_dyn}
    \dot{\b x} &= \b f_{\b x}(\b x) + \b g_{\b x}(\b x)v
\end{align}
with state $\b x \in \R^n$, input $v \in \R$, and functions $\b f_{\b x}: \R^n \to \R^n$ and $\b g_{\b x}: \R^n \to \R^n$ assumed to be continuously differentiable on $\R^n$. When analyzing underactuated systems, it will be useful to consider how actuation enters the system dynamics; to this end, consider an output $y: \R^n \to \R$. 
In order for the evolution of this output to be impacted by a controller, the input $v$ must appear in a derivative of $y$ in a meaningful way. Consider a time derivative of the output:
\begin{align*}
    \dot {y}(\b x) = \underbrace{\frac{\partial y}{\partial \b x}\b f_{\b x}(\b x)}_{L_{\b f_{\b x}} y(\b x)} + \underbrace{\frac{\partial y}{\partial \b x}\b g_{\b x}(\b x)}_{L_{\b g_{\b x}} y(\b x)} v 
\end{align*}
where $L_{\b f_{\b x}} y:\R^n \to \R$ and $L_{\b g_{\b x}} y:\R^n \to \R$ are the \textit{Lie derivatives} of the output $y$ with respect to the vector fields $\b f_{\b x}$ and $\b g_{\b x}$, respectively. If $L_{\b g_{\b x}} y(\b x) \equiv 0$, we can attempt to continue differentiating until a higher derivative is nonzero:
\begin{align*}
    {y}^{(\gamma)}(\b x) &= L_{\b f_{\b x}}^\gamma y(\b x) + L_{\b g_{\b x}}L_{\b f_{\b x}}^{\gamma-1}(\b x) v.
\end{align*}
Differentiating the output until the input appears is captured in the following notion of strict relative degree:

\begin{definition}
    \cite{sastry_linearization_1999} \label{def:rel_deg}
    An output $y :\R^n \to \R$ for the system \eqref{eqn:x_affine_dyn} is said to have relative degree $\gamma\in\mathbb{Z}_{+}$ at $\b x_0$ if:
    \begin{align*}
        L_{\b g_{\b x}}L_{\b f_{\b x}}^k(\b x) \equiv 0,\ \ \ \ 0\le k\le \gamma-2
    \end{align*}
    and $L_{\b g_{\b x}}L_{\b f_{\b x}}^{\gamma-1}(\b x) \neq 0$.
\end{definition}

Given an output of relative degree $\gamma \in \mathbb{Z}_+$, consider the mapping $\b \Phi_{\b \eta}:\R^n\to N\triangleq\R^\gamma$, defined as:
\begin{align}
    \b \Phi_{\b \eta}(\b x) \triangleq \begin{bmatrix}
        y(\b x) & \dot{y}(\b x) & \cdots & y^{(\gamma- 1)}(\b x)
    \end{bmatrix}^\top. \label{eqn:output}
\end{align}
We will subsequently take $\b \eta = \b\Phi_{\b \eta}(\b x)\in N$ to represent coordinates of the output space. Valid relative degree allows the constructive synthesis of controllers which exponentially stabilizes the outputs. 

\begin{definition}
    The signal $\b \eta(t)$ is \textit{exponentially stable} on domain $D \subset N$ if there exists $M, \lambda > 0$ such that:
\begin{align*}
 \b \eta_0 \in D \quad \implies \quad \|\b \eta(t)\| \leq M e^{-\lambda t} \|\b \eta_0\|.
\end{align*}
\end{definition}

Lyapunov theory \cite{sontag1989universal} states that exponential stability is one-to-one with the existence of a control Lyapunov function (CLF)
$V: N \to \R$ satisfying:
\begin{align}
    k_1\|\b \eta\|^2 \leq V(\b \eta) &\le k_2 \|\b \eta\|^2 \nonumber\\
    \inf_v \dot{V}(\b x, v) &\le -k_3 V(\b \eta). \label{eqn:lyap}
\end{align}
for $k_i > 0$. We define $\mathcal{K} = \{ k(\b x) ~\vert~ \dot{V}(\b x, k(\b x)) \leq -k_3 V(\b \eta)\}$ to be the set of all output exponentially stabilizing feedback controllers, which is nonempty under valid relative degree \cite{ames2014rapidly}. A common technique to stabilize outputs with valid relative degree is feedback linearization:
\begin{align*}
    k_{\text{fbl}}(\b x) = \left(L_{\b g_{\b x}}L_{\b f_{\b x}}^{\gamma-1}(\b x)\right)^{-1}(-L_{\b f_{\b x}}^\gamma(\b x) + u)
\end{align*}
for $k_{\text{fbl}}: \R^n \times \R \to \R$, with $u \in \R$ the auxiliary input. Under this controller, the $\b \eta$ dynamics become:
\begin{align*}
    \dot{\b \eta} = \underbrace{\begin{bmatrix} \b 0 & \b I \\ \b 0 & \b 0 \end{bmatrix}}_{\triangleq \b F}\b \eta + \underbrace{\begin{bmatrix}\b 0 \\ 1\end{bmatrix}}_{\triangleq \b G} u,
\end{align*}
for $\b F \in \R^{\gamma \times \gamma}, \b G \in \R^\gamma$. Once a system's available outputs are zeroed, the remaining states evolve on a manifold \cite{isidori_elementary_1995}, motivating the following discussion.

Consider a differentiable function $\b h:\R^n \to \R^p$ with $\b 0$ a regular value, i.e. $\frac{\partial\b h}{\partial \b x}\vert_{\b x = 0}$ is full rank. Then, we have that $\mathcal{M}\triangleq \{\b x~|~\b h(\b x) = \b 0 \}$ defines a $p$-dimensional embedded submanifold of $\R^n$ \cite{spivak_calculus_2018}. Associated with such a manifold is the notion of a tangent space. As the manifolds being considered are 0-level sets of a function $\b h$, a vector $\b v\in \R^n$ is a \textit{tangent vector} to a manifold $\mathcal{M}$ at the point $\b x \in \mathcal{M}$, denoted as $\b v\in \mathsf{T}_{\b x}\mathcal{M}$, if:   
    \begin{align*}
        \frac{\partial \b h}{\partial \b x}^\top \b v = \b 0.
    \end{align*}

This aligns with the classical notions of tangent vectors, as the gradient field of a function provides a basis for the annihilator of the \textit{tangent space}, the span of all tangent vectors at a point, of the manifold defined by a level set.

A key property will be the notion of controlled invariance for such a manifold, defined as:
\begin{definition}
    A manifold $\mathcal{M}$ is \textit{controlled invariant} under the dynamics \eqref{eqn:x_affine_dyn} if for all $\b x \in \R^n$ there exists an input $v\in \R$ such that:
    \begin{equation*}
        \b f_{\b x}(\b x) + \b g_{\b x}(\b x) v \in \mathsf T_{\b x} \mathcal{M}.
    \end{equation*}
\end{definition}
There must exist and input such that the vector field associated with the dynamics must lie in the tangent space of the manifold. Our proposed method aims to find controlled invariant manifolds with exponentially stable dynamics. 

Finally, we introduce the \textit{actuation decomposition}, which highlights the structure of actuated and unactuated states. When $y$ is valid relative degree, each $y^{(i)}$, for $i = 0, \hdots, \gamma -1$ are linearly independent, and $\b \eta$ forms a basis for $\gamma$ dimensions of $\R^n$ \cite{isidori_elementary_1995}. We can construct a set of normal coordinates $\b z \in Z \subset \R^{\nz}$, where $\nz = n - \nn$, via $\b \Phi_{\b z}: \R^n \to Z$. This transform is defined such that $\b \Phi: \R^n \to N \times Z$ given by:
\begin{align*}
    \begin{bmatrix}
        \b \eta \\ \b z
    \end{bmatrix} = \begin{bmatrix}
        \b \Phi_{\b \eta}(\b x) \\ \b \Phi_{\b z}(\b x)
    \end{bmatrix} \triangleq \b \Phi(\b x)
\end{align*}
is a diffeomorphism, and $\frac{\partial\b \Phi_{\b z}}{\partial\b x} \b g_{\b x} \equiv 0$ \cite{isidori_elementary_1995}. This implies that the $\b z$ dynamics are independent of the control input:
\begin{align} \label{eqn:act_decomp}
\begin{bmatrix}
    \dot{\b \eta} \\ \dot{\b z}
\end{bmatrix} = \underbrace{\begin{bmatrix}
    \b F \b \eta \\ \b \omega(\b \eta, \b z)
\end{bmatrix}}_{\b f(\b \zeta)} + \underbrace{\begin{bmatrix}
    \b G \\ \b 0
\end{bmatrix}}_{\b g(\b \zeta)} u
\end{align}
where $\b \zeta \triangleq (\b \eta, \b z) \in \mathcal{X} \triangleq N\times Z$. We deem this  transformation an \textit{actuation decomposition}; it separates the system states into a set of directly actuated coordinates ($\b \eta$), and a set of unactuated coordinates ($\b z$). As this decomposition takes a central role in our approach to underactuated control, \eqref{eqn:act_decomp} will be the starting point for dynamics in this paper. 

\section{Zero Dynamics Policies}

We propose a differentiable mapping $\b\psi: Z \rightarrow N$, which maps from the underactuated states $\b z$ to desired locations for the actuated states $\b \eta$. This is motivated by several results in robotics, such as the Raibert Heuristic, which maps a walking robot's center of mass (unactuated) to desired foot positions (actuated). The mapping $\b \psi$ induces a $\nz$ dimensional submanifold of $\mathcal{X}$ via the zero level set of the function $\b h(\b \eta, \b z) \triangleq \b \eta - \b \psi(\b z)$:
\begin{equation} \label{eq:zdp_manifold}
    \mathcal{M}_{\b \psi} \triangleq \{ (\b \eta, \b z) \in N \times Z \ \vert \ \b h(\b\eta, \b z) = \b 0\},
\end{equation}
as $\frac{\partial \b h}{\partial \b \zeta} = [\b I~ \frac{\partial \b \psi}{\partial \b z}]$ is full row rank. With this, we can now introduce the notion of \textit{zero dynamics}:

\begin{definition}
    The \textit{zero dynamics} associated with a controlled invariant manifold $\mathcal{M}_{\b \psi}$ are given by:
    \begin{align*}
        \dot{\b z} = \b\omega(\b\psi(\b z), \b z).
    \end{align*}
\end{definition}

\subsection{Required Properties of Zero Dynamics Policies}

If the zero dynamics are exponentially stable, we will show that stabilizing $\mathcal{M}_{\b \psi}$ stabilizes the whole system. To this end, we propose the following output:
\begin{align}\label{eqn:out}
    y = \eta_1 - \psi_1(\b z),
\end{align}
where $(\cdot)_i$ denotes the $i^{\textrm{th}}$ index. The following assumption is required to ensure $y$ can maintain relative degree $\nn$:
\begin{assumption}
    \label{ass:rel_deg}
    We have that $\frac{\partial \b \omega}{\partial \eta_i} = \b 0$ for all $i = 3,\ldots,\nn$.
\end{assumption}
This assumption is trivially satisfied for $\nn \leq 2$, a case typical for robotic systems. We now give a condition for the relative degree of \eqref{eqn:out}:
\begin{lemma} \label{lem:zdp_invariance}
    The output $y(\b \zeta) = \eta_1 - \psi_1(\b z)$ has valid relative degree $\gamma$ if and only if $\frac{\partial\psi_1}{\partial z}\frac{\partial \b\omega}{\partial \eta_2} \ne 1$.
\end{lemma}
\begin{proof}
    Taking derivatives until the input appears yields:
    \begin{align}
        \dot{y} &= \eta_2 - \frac{\partial\psi_1}{\partial \b z} \b \omega(\b \eta, \b z)\nonumber \\
        y^{(i)} &= \left(1 - \frac{\partial\psi_1}{\partial \b z} \frac{\partial \b \omega}{\partial \eta_2} \right) \eta_{i+1} + W_i(\b \eta_{1:i}, \b z), \label{eqn:output_dot}\\ 
        y^{(\gamma)} &= \left(1 - \frac{\partial\psi_1}{\partial \b z} \frac{\partial \b \omega}{\partial \eta_2} \right) u + W_{\gamma}(\b \eta, \b z)\nonumber 
    \end{align}
    for $i=2,\ldots,\gamma-1$ and where $W_i: N \times Z \to \R$ is introduced to hold additional terms for each derivative
    , and $\b \eta_{i:j} = [\eta_i \ \ldots \ \eta_j]^\top$.
    As $u$ does not appear until $y^{(\gamma)}$, we have that $L_{\b g} L_{\b f}^i y \equiv 0$ for $i = 0, \ldots, \gamma - 2$, and $L_{\b g} L_{\b f}^i y = 1 - \frac{\partial\psi_1}{\partial \b z} \frac{\partial \b \omega}{\partial \eta_2}$. Therefore, the output is relative degree $\gamma$ if and only if this term is nonzero. 
\end{proof}
\vspace{2mm}
In the case $\nn = 1$, the output \eqref{eqn:out} has valid relative degree one. 
Importantly, if $y$ is valid relative degree, each $y^{(i)}$, for $i = 0, \hdots, \gamma -1$ are linearly independent, and form a basis for $\gamma$ dimensions of $\mathcal{X}$ \cite{isidori_elementary_1995}. Defining the error coordinates:
\begin{align} \label{eqn:error}
    \b e = \begin{bmatrix}
        y & \dot{y} & \cdots & y^{(\gamma - 1)}
    \end{bmatrix}^\top \in \mathcal{E}
\end{align}
as $y$ and its first $\gamma - 1$ derivatives, we can construct the associated zeroing manifold of the output $y = \eta_1 - \psi_1(\b z)$.
\begin{lemma} \label{lem:output_zeroing}
Consider a controlled invariant manifold $\mathcal{M}_{\b\psi}$ and its associated output \eqref{eqn:out}. If the output has relative degree, then $\mathcal{M}_{\b\psi}$ is the zeroing manifold associated \eqref{eqn:out}.
\end{lemma}
\begin{proof}
Valid relative degree implies the existence of a unique zeroing manifold $\mathcal{M}$, an $\nz$ dimensional surface on which $\b e \equiv 0$ \cite{isidori_elementary_1995}. Because $\mathcal{M}_{\b \psi}$ is controlled invariant with $y \equiv 0$ (implying derivatives of $y$ are zero), it is also an $\nz$ dimensional zeroing surface; by uniqueness, $\mathcal{M} = \mathcal{M}_{\b \psi}$.
\end{proof}
\vspace{2mm}
We have shown that a controlled invariant manifold $\mathcal{M}_{\b \psi}$ is the zeroing manifold associated with \eqref{eqn:out} when this output is valid relative degree.
If a function $\b \psi$ can be found such that the zero dynamics on $\mathcal{M}_{\b \psi}$ are also stable, then the system can be constructively  stabilized as follows:
\begin{theorem} \label{thm:zdp_comp_lyap}
    Consider a relative degree $\gamma$ output  $y = \eta_1 - \psi_1(\b z)$ with zeroing manifold $\mathcal{M}_{\b \psi}$. If the zero dynamics of $\mathcal{M}_{\b \psi}$ are exponentially stable, then any output stabilizing controller $\b k \in \mathcal{K}$ renders the full state $\b \zeta$ exponentially stable.
\end{theorem}
\begin{proof}
    By Lemma \ref{lem:output_zeroing}, the relative degree of $y$ implies that holding $\b e = 0$ renders $\mathcal{M}_{\b \psi}$ invariant. Since $\mathcal{M}_{\b \psi}$ has exponentially stable dynamics when rendered invariant converse Lyapunov guarantees the existence of $V_{\b z}(\b z)$ satisfying:
    \begin{align*}
        k_{1,z} \|\b z\|^2 \leq V_{\b z}(\b z) &\leq k_{2,z} \|\b z\|^2 \\
        \dot{V}_{\b z}(\b z) = \frac{\partial V_{\b z}}{\partial \b z} \b \omega(\b \psi(\b z), \b z) &\leq -k_{3,z} \|\b z\|^2 \\
        \left\| \frac{\partial V_{\b z}}{\partial\b z} \right\| &\leq k_{4,z} \|\b z\|
    \end{align*}
    for $k_{i, z}>0$. Applying any controller $k(\b \zeta) \in \mathcal{K}$ (nonempty by virtue of valid relative degree) implies by converse Lyapunov the existence of a Lyapunov function on the error. Define $V_{\b e}(\b e)$ satisfying:
    \begin{align*}
        k_{1,e} \|\b e\|^2 \leq V_{\b e}(\b e) &\leq k_{2,e} \|\b e\|^2 \\
        \dot{V}_{\b e}(\b e) &\leq -k_{3,e} \|\b e\|^2
    \end{align*}
    for $k_{i, e} > 0$. We then use the implicit function theorem to establish that $\b \eta$ can be written as a function of $\b e, \b z$:
    \begin{align*}
        \b \xi(\b e, \b \eta, \b z) \triangleq \b e - \begin{bmatrix}
        y & \dot{y} & \hdots & y^{(\gamma-1)}
    \end{bmatrix}^\top = \b 0
    \end{align*}
    Observe from \eqref{eqn:output_dot} that $\frac{\partial\b \xi}{\partial \b \eta}$ is lower triangular. Furthermore, $(\frac{\partial\b \xi}{\partial \b \eta})_{1,1} = 1$ and $(\frac{\partial\b \xi}{\partial \b \eta})_{i,i} = (1 - \frac{\partial\psi_1}{\partial\b z} \frac{\partial\b \omega}{\partial \eta_2})$ for $i=2,\ldots,\gamma$.
    By assumption of relative degree, the diagonal elements are nonzero and therefore $\frac{\partial\b \xi}{\partial \b \eta}$ is invertible.
    Therefore, there exists a function $\b \Gamma: \mathcal{E} \times Z \to N$ such that $\b \xi(\b e, \b \Gamma(\b e, \b z), \b z) = 0$. We can then redefine the $\b z$ dynamics in terms of $\b e$ via:
    \begin{equation*}
        \Tilde{\b \omega}(\b e, \b z) = \b \omega\left(\b \Gamma(\b e, \b z), \b z\right) = \b \omega(\b \eta, \b z).
    \end{equation*}
    Finally, consider the positive definite function $V(\b e, \b z) = \sigma V_{\b e}(\b e) + V_{\b z}(\b z)$, for $\sigma > 0$, whose time derivative is:
    \begin{align}
        \dot{V}(\b e, \b z) &= \sigma \dot{V}_{\b e}(\b e) + \frac{\partial V_{\b z}}{\partial\b z}\Tilde{\b \omega}(\b e, \b z) \nonumber \\
        &= \dot{V}_{\b e}(\b e) + \frac{\partial V_{\b z}}{\partial\b z}\Tilde{\b \omega}(\b 0, \b z) +\frac{\partial V_{\b z}}{\partial\b z}\left(\Tilde{\b \omega}(\b e, \b z) - \Tilde{\b \omega}(\b 0, \b z)\right)\nonumber \\
        &\leq  -\sigma k_{3,e} \|\b e\|^2 - k_{3,z} \|\b z\|^2 + k_{4,z} L_\omega \|\b e\|\|\b z\| \nonumber\\
        &= -\begin{bmatrix}
            \|\b e\| \\ \|\b z\|
        \end{bmatrix}^\top \begin{bmatrix}
            \sigma k_{3,e} & -\frac{k_{4,z} L_\omega}{2} \\ -\frac{k_{4,z} L_\omega}{2} & k_{3,z}
        \end{bmatrix}\begin{bmatrix}
            \|\b e\| \\ \|\b z\|
        \end{bmatrix} \label{eqn:quad_form}
    \end{align}
    where $L_\omega$ is a Lipshitz constant of $\Tilde{\b \omega}$. Choosing $\sigma > \frac{k_{4,z}^2 L_\omega^2}{4k_{3,e} k_{3,z}}$ renders the matrix positive definite, and the quadratic form \eqref{eqn:quad_form} can be bounded, for $\lambda > 0$:
    \begin{equation*}
        \dot{V}(\b e, \b z) \leq -\lambda V(\b e, \b z)
    \end{equation*}
    since $V$ can be bounded by quadratic functions, certifying that the $V$ is a Lyapunov function by \eqref{eqn:lyap}. Thus, the composite system is exponentially stable under any controller which exponentially stabilizes the outputs. 
\end{proof}

\subsection{Local Existence of Stabilizing Zero Dynamics Policies}
Around the origin, we demonstrate via construction that $\b \psi$ exists for any locally controllable nonlinear system. 
To achieve this, we identify a stable zeroing manifold for the linear system and leverage the relationship between a nonlinear system and its linearization to generate an output with stable zero dynamics. 
%
To this end, consider the linearization of \eqref{eqn:act_decomp} about the origin:
\begin{align}
    \begin{bmatrix}
        \dot{\eta}_1 \\ \dot{\b \eta}_{2:\gamma-1} \\ \dot{\eta}_{\gamma} \\ \dot{\b z}
    \end{bmatrix} &= \underbrace{\begin{bmatrix}
            0 & 1 & \b 0 & \b 0 \\ \b 0 & \b 0 & \b I & \b 0 \\ 0 & 0 & \b 0 & \b 0 \\ \b a_{\eta_1} & \b a_{\eta_2} & \b 0 & \b A_{\b z}
    \end{bmatrix}}_{\triangleq \b A} \begin{bmatrix}
        {\eta_1} \\ {\eta_2} \\ {\b \eta}_{3:\gamma} \\ {\b z}
    \end{bmatrix} + \underbrace{\begin{bmatrix}
            0 \\ \b 0 \\ 1 \\ \b 0
    \end{bmatrix}}_{\triangleq \b B} u \label{eq:zd_linearized}
\end{align}
with $\b a_{\eta_i} = \frac{\partial \b\omega}{\partial \eta_i}\in \R^{\nz}$ and $ \b A_{\b z} = \frac{\partial \b \omega}{\partial \b z} \in \R^{\nz \times \nz}$. First, we identify a manifold invariant under a stabilizing controller:
\begin{lemma} \label{lem:invsub_lin}
    There exists a nonempty set of controllers which stabilize \eqref{eq:zd_linearized} and induce an $\nz$ dimensional invariant subspace $\mathcal{S}$ such that for each $\b z$ there exists a unique $\b \eta$ such that $(\b \eta, \b z) \in \mathcal{S}$.
\end{lemma}
\begin{proof}
    By assumption the nonlinear system is locally controllable; therefore \eqref{eq:zd_linearized} is controllable \cite{isidori_elementary_1995}. As such, the system can be stabilized by pole placement. Let $u = -\b K \b \zeta$ be any controller which places the poles at unique locations on the negative real axis. Then the closed loop system, $\b A - \b B\b K$, will have $n$ unique, linearly independent eigenspaces \cite{murray_feedback_systems}. It is therefore possible to pick $\nz$ distinct eigenvectors such that the projection onto $Z$ spans the $Z$ subspace. Let $\b v_1, \hdots, \b v_{\nz} \in \R^{n}$ be these eigenvectors. Define
    \begin{equation*}
        \b S = \begin{bmatrix}
        \b v_1 & \hdots & \b v_{\nz}
    \end{bmatrix}
    \end{equation*}
    Then $\mathcal{S} = \text{span}(\b S)$ is an invariant subspace of the closed loop dynamics, since it is the span of eigenspaces. 
    Given a point $\b z$, the corresponding point on $\mathcal{S}$ is given by 
    $$\begin{bmatrix}
        \b \eta \\ \b z
    \end{bmatrix} = \b S \left( \begin{bmatrix}
        0 & \b I
    \end{bmatrix} \b S\right)^{-1}\b z = \begin{bmatrix}
        \b S_{\b \eta}^\top \\ \b I
    \end{bmatrix}\b z$$
    where $\b S_{\b \eta} = \begin{bmatrix}
        \b s_{\eta_1} & \hdots & \b s_{\eta_\gamma}
    \end{bmatrix}$, with $\b s_{\eta_i} \in \R^{\nz}$. The matrix inverse is well defined since $\b S$ is selected such that its projection onto $\b z$ coordinates spans $Z$. 
\end{proof}
\begin{remark}
    While Lemma \ref{lem:invsub_lin} is proven for a specific form of controller (pole placement), nearly all stabilizing controllers will have $\nz$ dimensional invariant subspaces which are parameterizable by $\b z$.
    Any such subspace can be chosen, and the resulting $\b S$ can be used in the following analysis. 
\end{remark}

\cref{lem:invsub_lin} defines an invariant manifold for the controlled linear system. In order to appeal to composite stability, there must exist a controller rendering $\mathcal{S}$ attractive. This can be achieved by constructing an associated output with valid relative degree. 
However, from \cite{isidori_elementary_1995}, we know that there may exist zeroing manifolds that do not have valid relative degree. We now show that under the assumption of controllability, $\mathcal{S}$ is the zeroing manifold of a suitable output:


\begin{lemma} \label{lem:reldeg_lin}
    Consider a $\nz$ dimensional subspace $\mathcal{S}$ satisfying \cref{lem:invsub_lin}. The output $y = \eta_1 -\b s_{\eta_1}^\top \b z$ has valid relative degree $\gamma$ and $\mathcal{S}$ is the zeroing manifold for this output. 
\end{lemma}
\begin{proof}
    Take $\b C \triangleq \begin{bmatrix}
        1 & \b 0 & -\b s_{\eta_1}^\top
    \end{bmatrix}$ such that $y = \b C \b \zeta$. Consider the closed loop matrix $\b A_{cl} = \b A - \b B \b K$. For compactness, define:
    \begin{align*}
        m_k &= -\b s_{\eta_1}^\top \b A_{\b z}^k \b a_{\eta_1}\\
        \b O_k &= - \b s_{\eta_1}^\top \b A_{\b z} ^k \\
        q_k &= \b O_k \left( \b a_{\eta_1} + \b A_{\b z} \b a_{\eta_2}\right)\\
        p &= 1 - \b s_{\eta_1}^\top \b a_{\eta_2} 
    \end{align*}
    Using these we construct the matrix $\b E \in \R^{\gamma \times n}$:
    \begin{align}\label{eqn:linE}
    \setlength\arraycolsep{1.8pt}
        \b E = \begin{bmatrix}
            \b C \\ \b C \b A_{cl} \\ \b C\b A_{cl}^2 \\ \vdots \\ \b C \b A_{cl}^{\gamma - 1}
        \end{bmatrix} = \begin{bmatrix}
            1 & 0 & 0 & \cdots & 0 & \b O_0 \\
            m_0 & p & 0 & \cdots & 0 & \b O_1 \\
            m_1 & q_0 & p & \cdots & 0 & \b O_2 \\
            \vdots & \vdots & \vdots & \ddots & \vdots & \vdots \\
            m_{\gamma - 2} & q_{\gamma - 3} & q_{\gamma - 4} & \cdots & p & \b O_{\gamma - 1}
        \end{bmatrix}
    \end{align}
    Observe that for $i < \gamma - 1$, $\b C \b A_{cl} ^i \b B = 0$, as the $E_{j,\gamma} = 0$ for $j = 1, \ldots, \gamma - 1$. Thus, the output $y$ cannot be of relative degree less than $\gamma$. Assume for contradiction that the output is not relative degree $\gamma$. This implies: 
    \begin{equation}\label{eqn:contradiction}
        L_{\b B} L_{\b A_{cl}\b \zeta}^{\gamma - 1} y = \b C\b A_{cl}^{\gamma-1} \b B = p = 1 - \b s_{\eta_1}^\top \b a_{\eta_2} = 0.
    \end{equation}
    By \cref{lem:invsub_lin}, $\mathcal{S}$ is $\b A_{cl}$ invariant - this implies the existence of $\b J \in \R^{\nz \times \nz}$ such that $\b A_{cl} \b S = \b S \b J$, i.e. the image of $\b S$ under $\b A_{cl}$ is contained in $\b S$. As $\b C \b S = \b 0$, and $\b C \b A_{cl}^k \b S = \b C \b S \b J^k = 0$ we have that $\b S \in \text{ker}(\b E)$. Furthermore, by \eqref{eqn:contradiction} we have $\b C \b A_{cl}^{\gamma - 1} \b B = 0$, and $\b B \in \text{ker}(\b E)$. Finally, note that $\b B \notin \b S$, as $\b S$ contains an identity block in the lower $\nz$ rows, while $\b B$ has a corresponding $\b 0$ block. This implies $\text{dim}(\text{ker}(\b E)) \geq \text{dim}(\text{span}(\b S, \b B)) = \nz + 1$.
    By Rank Theorem, $\text{rank}(\b E) + \text{dim}(\text{ker}(\b E)) = n$, and therefore,
    \begin{align*}
        \text{rank}(\b E) \leq n - (\nz + 1) = \gamma - 1.
    \end{align*}
    Therefore, $\b E$ is rank deficient, since it has $\gamma$ rows. There is a nontrivial left nullspace; there exists $\b v = [\alpha_0 \ \hdots \ \alpha_{\gamma - 1}]^\top$, such that $\b v^\top \b E = \b 0$. Without loss of generality\footnote{$\b v$ must have at least one nonzero term since the null space is nontrivial. If $\alpha_{\gamma - 1} = 0$, this proof can be continued by redefining $\b E$ to omit the last row (or as many as necessary such that the last element of $\b v$ is nonzero). Then $\b v$ can be scaled such that its last entry has magnitude 1.}, we take $\alpha_{\gamma-1} = 1$. As the rightmost block of $\b v^\top \b E = \b 0$, 
   \begin{align}
       \b O_{\gamma - 1} &= -\sum_{i=0}^{\gamma - 2} \alpha_{i} \b O_i \label{eqn:crit}
   \end{align}    
    Next, because $p=0$ by \eqref{eqn:contradiction}, $\b v^\top \b E_{\gamma - 2} = \alpha_{\gamma -1} q_0 = 0$, where $\b E_{i}$ is the $i$'th column of $\b E$. Therefore, $q_0 = 0$. Applying this logic on rows of $\b E$ backward to $\b E_2$ gives:
    \begin{align*}
        q_{i} &= 0 \quad \forall i \in \{ 0, \hdots, \gamma - 3\}.
    \end{align*}
    Finally, we demonstrate via induction that $q_j = 0$, for any $j \in \mathbb{Z}_+$. Assuming that $q_j = 0$, we aim to show that $q_{j+1} =-\b s_{\eta_1}^\top \b A_{\b z}^{j+1}\left( \b a_{\eta_1} + \b A_{\b z} \b a_{\eta_2}\right)= 0$. Introduce $q_{j+1}$ by right-multiplying \eqref{eqn:crit} by $\b A_{\b z}^{k}(\b a_{\eta_1} + \b A_{\b z} \b a_{\eta_2})$, for $k = j+1-(\gamma - 1)$:
    \begin{align*}
        q_{j+1} &= \b O_{\gamma - 1}\b A_{\b z}^{k}(\b a_{\eta_1} + \b A_{\b z} \b a_{\eta_2}) \\
        &= -\sum_{i=0}^{\gamma - 2}\alpha_{i} \b s_{\eta_1}^\top \b A_{\b z}^{i + k} \b a_{\eta_1} - \sum_{i=0}^{\gamma - 2}\alpha_{i} \b s_{\eta_1}^\top \b A_{\b z}^{i + k + 1} \b a_{\eta_2}.
    \end{align*}
    By the induction hypothesis $q_i = 0$, for $i = 0, \hdots, j$ which can be rearranged:
    \begin{align*}
        -\b s_{\eta_1}^\top \b A_{\b z}^{i} \b a_{\eta_1} = \b s_{\eta_1}^\top \b A_{\b z}^{i+1}\b a_{\eta_2}.
    \end{align*}
    Substituting this for terms $q_{j-(\gamma - 1)}$ to $q_{j}$ into the right sum:
    \begin{align*}
        q_{j+1} &= -\sum_{i=0}^{\gamma - 2}\alpha_{i} \b s_{\eta_1}^\top \b A_{\b z}^{i + k} \b a_{\eta_1} + \sum_{i=0}^{\gamma - 2}\alpha_{i} \b s_{\eta_1}^\top \b A_{\b z}^{i + k } \b a_{\eta_1} = 0.
    \end{align*}
    And the inductive step has been shown. The base case holds trivially for $\gamma \geq 3$ by the structure of $\b E$, and $\b C \b B = 1$ for relative degree one systems (trivially valid relative degree). It remains to show the base case holds for $\gamma = 2$.  

    First, examine the fact that $\mathcal{S}$ is invariant under $\b A_{cl}$.
    Expanding $\b A_{\text{cl}} \b S = \b S \b J$, given that $\b K = [k_{\eta_1} \ k_{\eta_2} \ \b k_{\b z}]$:
    \begin{align*}
        \begin{bmatrix}
                \b s_{\eta_2}^\top \\-k_{\eta_1} \b s_{\eta_1}^\top - k_{\eta_2} \b s_{\eta_2}^\top - \b k_{\b z}
             \\ \b a_{\eta_1} \b s_{\eta_1}^\top + \b a_{\eta_2} \b s_{\eta_2}^\top + \b A_{\b z}
        \end{bmatrix} &= \begin{bmatrix}
                \b s_{\eta_1}^\top\b J \\ \b s_{\eta_2}^\top\b J \\ \b J
        \end{bmatrix}
    \end{align*}
    Right multiply the second equation by $\b a_{\eta_2}$, and using \eqref{eqn:contradiction}:
    \begin{align*}
        \b a_{\eta_1} + \b A_{\b z} \b a_{\eta_2} = (\b I - \b a_{\eta_2}\b s_{\eta_1} ) \b J \b a_{\eta_2}
    \end{align*}
    And note that left multiplying by $\b s_{\eta_1}$ gives:
    \begin{align*}
        q_0 = \b s_{\eta_1}^\top (\b a_{\eta_1} + \b A_{\b z} \b a_{\eta_2}) = \b s_{\eta_1}^\top(\b I - \b a_{\eta_2}\b s_{\eta_1}^\top ) \b J \b a_{\eta_2} = 0
    \end{align*}
    again leveraging $\b s_{\eta_1}^\top \b a_{\eta_2} = 1$. We have established the based case for induction, for relative degree $\gamma = 2$. 

    We now aim to demonstrate the contradiction, by showing that \eqref{eqn:contradiction} leads to the system losing controllability.
    Consider the controllability matrix of the system. Controllability is preserved under feedback, so we examine $\mathscr{C}(\b A, \b B)$: 
    \begin{align*}
        \mathscr{C}(\b A, \b B) &= \begin{bmatrix}
            \b 0  & 1 & 0 & \hdots & 0\\
            \b I  & 0 & 0 & \hdots & 0\\
            \b 0 & \b a_{\eta_2} & \b Q_0 & \hdots & \b Q_{n-(\gamma + 1)}
        \end{bmatrix} \\
        \b Q_i &= \b A_{\b z}^{i}(\b a_{\eta_1} + \b A_{\b z} \b a_{\eta_2})
    \end{align*}
    We finish by showing the $\b C$ is in the left null space of $\mathscr{C}$:
    \begin{align*}
        \b C \mathscr{C} = \begin{bmatrix}
            0 & p & q_0 & \hdots & q_{n-(\gamma + 1)}
        \end{bmatrix} = \b 0
    \end{align*}
    And therefore $\mathscr{C}$ is not full rank. We have reached contradiction, and $y$ must be relative degree $\gamma$. 
    Finally, note that $\mathcal{S}$ is the zeroing surface associated with $y$ by \cref{lem:output_zeroing}.
\end{proof}
\vspace{2mm}
\cref{lem:reldeg_lin} constructs an output and its associated zeroing manifold such that 1) the output has valid relative degree, and 2), the zeroing manifold is exponentially stable. Now, we show that the output locally retains both of these properties under the nonlinear dynamics:
\begin{theorem} \label{thm:local_existance}
   \textit{Given a nonlinear system \eqref{eqn:act_decomp}, the output $y = \eta_1 - \b s_{\eta_1}^\top\b z$ obtained via linearization in \cref{lem:reldeg_lin} has valid relative degree and exponentially stable zero dynamics for the nonlinear system. As such, stabilizing $\b e \rightarrow 0$ results in stability of the entire system in a neighborhood of the origin.}
\end{theorem}

\vspace{0.2cm}
\begin{proof}
    First, we establish that the output is relative degree $\gamma$ for the nonlinear system. The output has the form $y = \b C \b \zeta = \eta_1 - \b s_{\eta_1}^\top \b z$. By \cref{lem:zdp_invariance}, we have that $L_{\b g}L_{\b f}^jy(\b \zeta) \equiv 0$ for $j = 0,\hdots, \gamma-2$ and:
    \begin{align*}
        L_{{\b g}} L_{{\b f}}^{\gamma - 1} y &= 1 - \b s_{\eta_1}^\top \frac{\partial \b \omega}{\partial \eta_2}.
    \end{align*}
    For relative degree, we need $L_{{\b g}} L_{{\b f}}^{\gamma - 1} y \neq 0$. 
    \begin{align*}
        L_{{\b g}} L_{{\b f}}^{\gamma - 1} y &= 1 - \b s_{\eta_1}^\top \frac{\partial \b \omega}{\partial \eta_2}\\
        &= \underbrace{1 - \b s_{\eta_1}^\top \b a_{\eta_2}}_{\b C \b A^{\gamma - 1} \b B} + \underbrace{\b s_{\eta_1}^\top \left(\b a_{\eta_2} - \frac{\partial \b \omega}{\partial \eta_2} \right)}_{\triangleq \Delta(\b \zeta)}
    \end{align*}
    where $\Delta: \mathcal{X} \to \R$. \cref{lem:reldeg_lin} assures that $|\b C\b A^{\gamma - 1} \b B| \triangleq \delta > 0$. To guarantee $L_{{\b g}} L_{{\b f}}^{\gamma - 1} y \neq 0$, we bound $|\Delta(\b \zeta)| < \delta$. 
    \begin{align*}
        \left| \Delta(\b \zeta) \right| &\leq \|\b s_{\eta_1}\| \left\|\b a_{\eta_2} - \frac{\partial\b \omega}{\partial \eta_2}\right\|
    \end{align*}
    Note that the function $\left\|\b a_{\eta_2} - \frac{\partial\b \omega}{\partial \eta_2}\right\|$ is continuous and zero at the origin. Therefore, there exists an $\varepsilon > 0$ such that
    \begin{align*}
        \left\|\b a_{\eta_2} - \frac{\partial\b \omega}{\partial \eta_2}\right\| \leq \frac{\delta}{2 \|\b s_{\eta_1}\|} \quad \forall \ (\b \eta, \b z) \in B_\varepsilon(\b 0, \b 0)
    \end{align*}
    Inside this epsilon ball, we have $|\Delta(\b \zeta)| \leq \frac{1}{2}\delta$, and therefore $|L_{{\b g}}L_{{\b f}}^{\gamma - 1}y | > \frac{1}{2}\delta$. Locally, the output has valid relative degree for the nonlinear system. 

    It remains to show that the zeroing manifold, $\mathcal{M}_{\b \psi}$ of this output is stable for the nonlinear system. First, given the error coordinates \eqref{eqn:error}, consider $\frac{\partial \b e}{\partial \b \zeta} = \begin{bmatrix}
        \frac{\partial\b e}{\partial \b \eta} & \frac{\partial \b e}{\partial \b z}
    \end{bmatrix}$ evaluated at the origin. For a row of this matrix,
    \begin{align}
        \frac{\partial}{\partial \b \zeta}\b e_i \bigg\vert_{\b \zeta = \b 0} &= \frac{\partial}{\partial\b \zeta} L_{\b f}^i y\bigg\vert_{\b \zeta = \b 0} \nonumber \\
        &= \frac{\partial^2 L_{\b f}^{(i-1)} y}{\partial\b \zeta^2}\b f\bigg\vert_{\b \zeta = \b 0} + \frac{\partial}{\partial\b \zeta}L_{\b f}^{(i-1)} y\frac{\partial \b f}{\partial\b \zeta}\bigg\vert_{\b \zeta = \b 0}  \nonumber \\
        &= \frac{\partial}{\partial\b \zeta}L_{\b f}^{(i-1)} y\frac{\partial \b f}{\partial\b \zeta}\bigg\vert_{\b \zeta = \b 0} \label{eqn:e_induct}\\
        &= \b C \b A^i \nonumber
    \end{align}
    where the third lines holds by $\b f(\b 0) = \b 0$ and the last equality can be shown by induction. The base case holds as $\frac{\partial y}{\partial\b \zeta} = \b C$, and the induction step is given by \eqref{eqn:e_induct}. We have:
    \begin{align}
        \frac{\partial \b e}{\partial \b \zeta}\bigg\vert_{\b \zeta = \b 0} = \begin{bmatrix}
            \b C \\ \vdots \\ \b C \b A^{\gamma - 1}
        \end{bmatrix}
    \end{align}
    which is precisely the same as \eqref{eqn:linE}. Therefore, $\frac{\partial \b e}{\partial \b \eta}$ is lower triangular with nonzero elements on the diagonal, and is thus invertible. By the implicit function theorem, there exists $\b \psi: Z \to N$ such that $\b e(\b \psi(\b z), \b z) = \b 0$. We aim to show that $\mathcal{M}_{\b \psi}$ is stable. Consider that by the implicit function theorem:
    \begin{align}
        \frac{\partial \b \psi}{\partial \b z} = -\frac{\partial \b e}{\partial \b \eta}^{-1} \frac{\partial \b e}{\partial \b z}. \label{eq:dpsi_dz}
    \end{align}
    Furthermore, note that since $\frac{\partial\b e}{\partial \b \zeta}\vert_{\b \zeta = \b 0} = \b E$, \eqref{eq:dpsi_dz} also holds for the linearization. 
    Therefore, at the origin, the tangent space of $\mathcal{M}_{\b \psi}$ for the nonlinear system is equal to the invariant subspace used by the linearization to design the output, i.e. $\frac{\partial\b e}{\partial \b \eta}\vert_{\b \zeta = \b 0} = \b S_{\b \eta}$. On $\mathcal{M}_{\b \psi}$ we have $\b \eta = \b \psi(\b z)$, or equivalently:
    \begin{align*}
        \b \eta &= \frac{\partial\b \psi}{\partial\b z}\bigg\vert_{\b \zeta = \b 0}\b z + \left(\b \psi(\b z) - \frac{\partial\b \psi}{\partial\b z}\bigg\vert_{\b \zeta = \b 0}\b z\right) \\
        &= \b S_{\b \eta}\b z + \underbrace{\left(\b \psi(\b z) - \b S_{\b \eta}\b z\right)}_{\b \Gamma(\b z)}
    \end{align*}
    By mean value theorem, there exists $\varepsilon > 0$ such that inside a ball $B_\varepsilon(\b 0, \b 0)$ we have 
    \begin{equation*}
        \|\Gamma(\b z)\| \leq N_\Gamma(\varepsilon) \|\b z\|
    \end{equation*}
    With $N_\Gamma(\varepsilon) \rightarrow 0$ as $\varepsilon \rightarrow 0$. 

    Finally, given that the surface $\mathcal{S}$ is stable under the linearized dynamics from Lemma \ref{lem:invsub_lin}, converse Lyapunov guarantees the existence of a function $V_{\b z}(\b z)$ satisfying
    \begin{align*}
        k_1 \|\b z\|^2 \leq V_{\b z}(\b z) &\leq k_2\|\b z\|^2 \\
        \dot{V}_{\b z}(\b z) &= \frac{\partial V_{\b z}}{\partial\b z}\left(\b a_{\eta_1} \b s_{\eta_1} + \b a_{\eta_2} \b s_{\eta_2} + \b A_{\b z}\right)\b z \\&\leq -k_3 \|\b z\|^2 \\
        \left\|\frac{\partial V_{\b z}}{\partial\b z}\right\| &\leq k_4 \|\b z\|
    \end{align*}
    for $k_i > 0$. We aim to show that this function is also a Lyapunov function for the dynamics on the nonlinear zeroing manifold, where $(\b \eta, \b z) = (\b \psi(\b z), \b z)$. To this end, define $\Tilde{\b A} \in \R^{\nz \times \nz}$ and $\b \Delta_1, \b \Delta_2: Z \to \R^{\nz}$: 
    \begin{align*}
        \Tilde{\b A} &= \b a_{\eta_1} \b s_{\eta_1} + \b a_{\eta_2} \b s_{\eta_2} + \b A_{\b z} \\
        \b \Delta_1(\b z) &= \b \omega\left(\b S_{\b \eta} \b z, \b z\right) - \Tilde{\b A}\b z \\
        \b \Delta_2(\b z) &= \b \omega\left(\b \psi(\b z), \b z\right) - \b \omega\left(\b S_{\b \eta}, \b z\right)
    \end{align*}
    Differentiate $V_{\b z}$ under the nonlinear dynamics:
    \begin{align*}
        \dot{V}_{\b z} &= \frac{\partial V_{\b z}}{\partial\b z}\b \omega\left(\b \psi(\b z), \b z\right) \\
        &=\frac{\partial V_{\b z}}{\partial\b z}\Tilde{\b A}\b z + \frac{\partial V_{\b z}}{\partial\b z}\left(\b \Delta_1(\b z) + \b \Delta_2(\b z) \right)        
    \end{align*}
    Noting that $\b \Delta_1(\b z)$ is the error in the linearization of $\b \omega$ on the linear surface $\mathcal{S}$, it can be bounded using mean value theorem as $\|\b \Delta_1(\b z)\| \leq N_1(\varepsilon) \|\b z\|$. Similarly, $\b \Delta_2$ can be bounded using Lipschitz continuity of $\b \omega$, and the fact that $\b \psi(\b z) - \b S_{\b \eta}\b z = \b \Gamma(\b z)$ to obtain $\|\b \Delta_2(\b z)\| \leq L_{\b \omega} N_\Gamma(\varepsilon) \|\b z\| = N_2(\varepsilon)\|\b z\|$. Letting $N(\varepsilon) = N_1(\varepsilon) + N_2(\varepsilon)$, we have
    \begin{align*}
        \left\|\frac{\partial V_{\b z}}{\partial\b z}\left(\b \Delta_1(\b z) + \b \Delta_2(\b z) \right)\right\| &\leq \left\|\frac{\partial V_{\b z}}{\partial\b z}\right\|\left(\|\b \Delta_1(\b z)\| + \|\b \Delta_2(\b z)\| \right) \\
        &\leq k_4 N(\varepsilon) \|\b z\|^2
    \end{align*}
    Choosing $\varepsilon > 0$ such that $N(\varepsilon) < \frac{k_3}{2k_4}$, we have
    \begin{align*}
        \dot{V}_{\b z}(\b z) & \leq -k_3\|\b z\|^2 + k_4 N(\varepsilon)\|\b z\|^2 \\
        &\leq - \frac{k_3}{2}\|\b z\|^2
    \end{align*}
    Therefore, we see that $V_{\b z}$ is a Lyapunov function on the nonlinear zeroing manifold, with a sufficiently small ball around the origin. We have successfully demonstrated that $\b \psi$ has the required properties to apply Theorem \ref{thm:zdp_comp_lyap} and conclude local exponential stability of the composite system. 
\end{proof}
\vspace{2mm}

This proof uses linearization to design an output which has desirable properties. Locally, the zeroing manifold for this output $\mathcal{M}_{\b \psi}$ is close to the linear systems zeroing manifold (captured by $\b \Delta_2$), and the zero dynamics are close to the dynamics of the linear system (captured by $\b \Delta_1$). Sufficiently close to the origin, relative degree and stability of $\mathcal{M}_{\b \psi}$ are retained for the nonlinear system.

It is important to emphasize that \cref{thm:local_existance} gives a completely \emph{constructive method for stabilizing a broad class of underactuated systems using output stabilization}. In the results section, we demonstrate this constructive method on a canonical example of underactuation, the cartpole.

\section{Optimal Control for Learning Zero Dynamics Policies}

While the linearization of a system about equilibrium can be used to construct locally stabilizing controllers, it is desirable to obtain larger regions of attraction, and leverage the nonlinear dynamics of the system. Therefore, we aim to construct $\man$ satisfying controlled invariance and stability.

\subsection{Optimal Control for Stabilization}

As asymptotic stability is a necessary condition for optimality \cite{liberzon_calculus_2012}, we leverage optimal control to find $\man$. Consider the infinite-time optimal control problem:
\begin{align}
    V(\b \zeta_0) \triangleq \min_{\substack{\b \zeta,  \b u}}\quad & \int_0^\infty c(\b \zeta(t),  \b u(t)) dt \label{eqn:ocp}\\
\textrm{s.t.} \quad & \dot {\b \zeta} = {\b f}(\b \zeta) + \b g(\b \zeta) \b u \notag
\end{align}
where $V: \mathcal{X} \rightarrow \R$ is the value function and $c:\mathcal{X} \times \mathcal{U} \to \R$ is a positive definite cost function. In order to apply \cref{thm:zdp_comp_lyap}, we will require exponential stability on the manifold. Therefore, we begin by stating conditions under which the optimal controller is exponentially stabilizing:
\begin{theorem} \label{thm:ocp_stab}
    \textit{Let $V(\b \zeta)$ be the value function for the optimal control problem defined \eqref{eqn:ocp}, with quadratic cost $c(\b \zeta, \b u) = \b \zeta^\top \b Q \b \zeta + ru^2$, with $\b Q \in \R^{n \times n}$ positive definite, $r > 0$ and compact state space $\mathcal{X}$.  The nonlinear system is exponentially stable under the optimal controller. }
\end{theorem}
\begin{proof}
    In a sufficiently small ball around the origin, the LQR approximation of the optimal controller, obtained by linearizing the dynamics about equilibrium, will be exponentially stabilizing for the nonlinear system \cite{sastry_linearization_1999}, as it locally satisfies input bounds. This implies constants $M_{LQR}, \ \lambda_{LQR}, \ \delta > 0$ such that:
    \begin{align*}
    \|\b \zeta_0\| \leq \delta  \quad \implies \quad 
        \|\b \zeta(t)\| \leq M_{LQR}e^{-\lambda_{LQR} t} \|\b \zeta_0\|
    \end{align*}
    We aim to show that the trajectory emanating from an arbitrary initial condition $\b \zeta_0 \in B_\delta(\b 0)$ is exponentially stable. For any $M,\lambda > 0$, consider the set:
    \begin{align*}
        T = \{ t \geq 0 ~\vert~ \|\b \zeta(t)\| > Me^{-\lambda t} \|\b \zeta_0\| \}
    \end{align*}
    We condition on whether there is an upper bound to the elements of $T$:
    
    \proofsubstep{Case 1:} There exists an upper bound $\overline{T}$ such that $t < \overline{T}$ for all $t \in T$. Then consider the maximum violation ratio
    \begin{align*}
        \overline{r} = \sup_{t \in T} \frac{\|\b \zeta(t)\|}{Me^{-\lambda t} \|\b \zeta_0\|} \leq \frac{B}{Me^{-\lambda \overline{T}}\|\b \zeta_0\|} 
    \end{align*}
    Take $\overline{r} = 1$ if $T$ is empty. Then:
    \begin{align*}
        \|\b \zeta(t)\| \leq \overline{r}Me^{-\lambda t} \|\b \zeta_0\|
    \end{align*}
    implying the trajectory is exponentially stable.
    
    \proofsubstep{Case 2:} There is no upper bound on the elements in $T$. We will establish $V(\b \zeta)$ is a Lyapunov function certifying exponential stability of the trajectory. Bound the decrease:
    \begin{align}
        \dot{V}(\b \zeta) &= -\b \left(\b \zeta^\top \b Q \b \zeta + ru^2 \right) \nonumber \\
        &\leq -\underline{\lambda}(\b Q) \|\b \zeta\|^2 \label{eq:value_dot}
    \end{align}
    Next, bound $V$ above by a quadratic function. Because LQR is suboptimal for the nonlinear system, applying it can only increase the cost relative to $V(\b \zeta)$:
    
    \begin{align*}
        V(\b \zeta_0) &\leq \int_0^\infty \b \zeta^\top \b Q \b \zeta + r(\b K \b \zeta)^2 dt \\
        &\leq \int_0^\infty \b (\overline{\lambda}(\b Q) + r \overline{\lambda}(\b K^\top \b K)) \|\b \zeta\|^2 dt\\
        &\leq \int_0^\infty \b (\overline{\lambda}(\b Q) + r \overline{\lambda}(\b K^\top \b K)) M_{LQR}^2 e^{-2\lambda t}\|\b \zeta_0\|^2 dt\\
        &= \frac{(\overline{\lambda}(\b Q) + r \overline{\lambda}(\b K^\top \b K)) M_{LQR}^2}{2\lambda_{LQR}} \|\b \zeta_0\|^2
    \end{align*}
    with $\overline{\lambda}, \underline{\lambda}$ the maximum and minimum eigenvalues respectively. Finally, lower bound $V(\b \zeta)$ by a quadratic.
    \begin{align*}
        V(\b \zeta_0) &= \int_0^\infty \b \zeta^\top \b Q \b \zeta + ru^2 dt \\
        &\geq \int_T \underline{\lambda}(\b Q) \|\b \zeta\|^2 dt\\
        &\geq \underline{\lambda}(\b Q) M^2 \|\b \zeta_0\|^2 \left(\int_T e^{-2\lambda t}  dt \right) \\
        &= \underline{\lambda}(\b Q) M^2 \|\b \zeta_0\|^2 \left(\int_0^\infty e^{-2\lambda t}  dt - \int_{\R_{\geq 0} \backslash T} e^{-2\lambda t}  dt\right) \\
        &= \underline{\lambda}(\b Q) M^2\left(\frac{1}{2\lambda} -c \right)\|\b \zeta_0\|^2
    \end{align*}
    where $\R_{\geq 0} \backslash T$ is the set difference between the nonnegative reals and $T$, and $\frac{1}{2\lambda} -c > 0$ as both integrals integrate over the same strictly positive function, but the right integral does so over a smaller domain.
    The bounds hold at each point on the trajectory, and $V$ is a Lyapunov function certifying exponential stability of the trajectory.

    We now extend this claim over the compact state space $\mathcal{X}$. At $V \succ 0$ and $\dot{V} \prec 0$, we have that the optimal controller is asymptotically stabilizing \cite{liberzon_calculus_2012}. By compactness of $\mathcal{X}$ and \eqref{eq:value_dot}, the time for a trajectory to enter $B_\delta(\b 0)$ is bounded by:
    \begin{align*}
        T_{\max} = \frac{\sup_{\b \zeta_0 \in \mathcal{X}} V(\b \zeta_0)}{\inf_{\b \zeta_0 \in \mathcal{X} \backslash B_\delta(\b 0)} \dot{V}(\b \zeta_0)} \leq  \frac{\sup_{\b \zeta_0 \in \mathcal{X}} V(\b \zeta_0)}{\underline{\lambda}(\b Q) \delta^2}
    \end{align*}
    Because trajectories in $B_\delta(\b 0)$ converge exponentially:
    \begin{align*}
        \|\b \zeta(t)\| \leq Me^{-\lambda (t - T_{\max})} \|\b \zeta(T_{\max})\| \quad \forall t > T_{\max}
    \end{align*}
    By compactness of $\mathcal{X}$, trajectories are bounded by $\|\b \zeta\| \leq B$, and the whole trajectory can be bounded exponentially:
    \begin{align*}
        \|\b \zeta(t)\| \leq \frac{\max\{M, B\} e^{\lambda T_{\max}}}{\min\{1, \delta\}}e^{-\lambda t} \|\b \zeta_0\| 
    \end{align*}
    The optimal controller is exponentially stabilizing.
\end{proof}
\vspace{2mm}

\subsection{Learning Zero Dynamics Policies}
\begin{figure*}
    \centering
    \includegraphics[width=\textwidth]{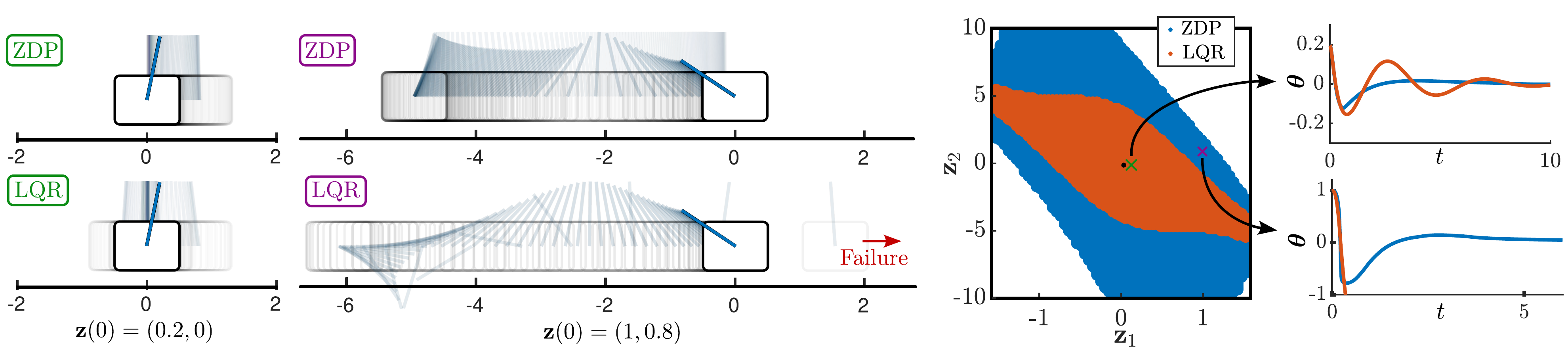}
    \caption{Zero Dynamics Policies (ZDPs) compared to LQR for the nonlinearly damped cartpole. Left: Simulated initial conditions for the cartpole, where LQR has slow response times and instabilities. Right: Region of attraction ($x=\dot{x}=0$) for the two methods, as well as pendulum angle over time.}
    \label{fig:theory}
    \vspace{-4mm}
\end{figure*}

We propose learning $\b\psi $ so $\mathcal{M}_{\b \psi}$ is controlled invariant and stable. Let $u^*(\b \zeta)$ solve the optimal control problem \eqref{eqn:ocp}. Then $\mathcal{M}_{\b \psi}$ is invariant under $u^*(\b \zeta)$ if for $\b \zeta_{\b \psi} = (\b \psi(\b z), \b z)$:
\begin{align*}
    \begin{bmatrix}
        \b I & -\frac{\partial \b \psi}{\partial\b z}
    \end{bmatrix} \begin{bmatrix}
        \hat{\b f}(\b \zeta_{\b \psi}) + \hat{\b g}(\b \zeta_{\b \psi}) u^*(\b \zeta_{\b \psi}) \\
        \b \omega(\b \zeta_{\b \psi})
    \end{bmatrix} = \b 0
\end{align*}
Given a neural network parameterization of $\b \psi_{\b \theta}$, we define the loss function:

\begin{align} \label{eqn:loss}
    \mathcal{L}(\b \theta) = \mathop{\mathbb{E}}_{\b z \sim D} \left\|\hat{\b f}(\b \zeta_{\b \theta}) + \hat{\b g}(\b \zeta_{\b \theta}) u^*(\b \zeta_{\b \theta}) - \frac{\partial \b \psi_{\b \theta}}{\partial\b z}\b \omega(\b \zeta_{\b \theta}) \right\|
\end{align}
for $\b \zeta_{\b \theta} = (\b \psi_{\b \theta}(\b z), \b z)$. Zero loss implies invariance of $\mathcal{M}_{\b \psi}$ under the optimal control which gives stability, by \cref{thm:ocp_stab}.
We minimize this loss using stochastic gradient descent.


\subsection{Application to the Cartpole}

We deploy the ZDPs on a classic underactuated system: the cartpole. We add nonlinear damping to the base coordinates of the form $d(\dot x) = \sigma(\dot x)\dot x$ where $\sigma(\dot x) = 0$ if $|\dot x| < 1e^{-3}$ and $1$ otherwise. This helps explore the effect of nonlinearities on the degradation of LQR performance, and how our nonlinear method compares. We take $\b Q = \b I$ and $\b r = 0.01$.
The ZDP policy was trained in the JAX module using iLQR to approximate $u^*$ and its gradient for training. A 2 layer, 256 neuron feedforward neural network with ReLU activations was pretrained with LQR and then minimized \eqref{eqn:loss}. 
We stabilized $\man$ with a PD controller and a feedforward term.
Our code can be found at \cite{code}.

Observe the performance of LQR versus ZDPs in Figure~\ref{fig:theory}. 
Even initial conditions close to the origin and within the domain of attraction of LQR, the modified cartpole's unstable nonlinear damping significantly slowed the controller's by inducing oscillations.
In comparison, ZDPs have smoother behavior and a larger region of attraction.

\section{Conclusion}
We proposed a method of constructing feedback controllers for underactuated systems. 
We split the controller design process into two steps: 1) learning a manifold that is invariant under optimal control and 2) applying output feedback to this manifold.
We proved that such a manifold exists for a broad class of nonlinear systems, and demonstrated the effectiveness of the ZDP method towards stabilizing the canonical underactuated system of the cartpole.
Future work includes proving the existence of such a manifold over a larger domain, and applying this method to hardware. 

\bibliographystyle{IEEEtran}
\balance
\bibliography{main.bib}

\end{document}